\documentclass[prc,twocolumn,floatfix,groupedaddress,nofootinbib,preprintnumbers,
amsmath,amssymb,amsfonts,superscriptaddress,widetable] {revtex4}
\usepackage{bm}
\usepackage{mathrsfs}
\usepackage{amssymb}
\usepackage{amsmath}
\usepackage{graphicx}
\usepackage{array}
\usepackage{color}
\usepackage{tabularx}
\newcommand{\Msun}{\,M_{\odot}}

\begin{document}
\title{Large Sound Speed in Dense Matter and the Deformability of Neutron Stars}


\author{Brendan Reed}\email{reedbr@iu.edu}
\affiliation{Department of Astronomy, Indiana University,
                  Bloomington, IN 47405, USA}
\affiliation{Center for Exploration of Energy and Matter and
                  Department of Physics, Indiana University,
                  Bloomington, IN 47405, USA}
\author{C. J. Horowitz}\email{horowit@indiana.edu}
\affiliation{Center for Exploration of Energy and Matter and
                  Department of Physics, Indiana University,
                  Bloomington, IN 47405, USA}
\date{\today}
\begin{abstract}
The historic first detection of the binary neutron star merger GW170817 by the LIGO-Virgo collaboration has set a limit on the gravitational deformability of neutron stars. In contrast, radio observations of PSR J0740+6620 find a very massive neutron star. Tension between the small deformability and the large maximum mass may suggest that the pressure rises rapidly with density and thus the speed of sound in dense matter is likely a large fraction of the speed of light.  We use these observations and simple constant sound-speed model equations of state to set a lower bound on the maximum speed of sound in neutron stars. 
If the tidal deformability of a 1.4$\Msun$ neutron star is less than 600, as is suggested by subsequent analyses of GW170817, we find that the sound speed in the cores of neutron stars is likely larger than the conformal limit of $c/\sqrt{3}$.  Implications of this for our understanding of both hadronic and quark-gluon descriptions of dense matter are discussed.  


\end{abstract}

\maketitle
\section{Introduction}
The historic gravitational wave observations of the binary neutron star (BNS) merger GW170817 impose significant constraints on the equation of state (EOS) of dense matter and limit the tidal polarizability $\Lambda$.  The \emph{tidal polarizability} (or deformability) is an intrinsic neutron-star (NS) property highly sensitive to the stellar compactness\,\cite{Hinderer:2007mb,Hinderer:2009ca,Damour:2009vw,
Postnikov:2010yn,Fattoyev:2012uu,Steiner:2014pda} that describes the tendency of a NS to develop a mass quadrupole as a response to the tidal field induced by its companion\,\cite{Damour:1991yw,Flanagan:2007ix}. The dimensionless tidal polarizability $\Lambda$ is defined as follows:
 $\Lambda = \frac{2}{3}k_{2} (c^{2}R/ GM)^{5}$
where $k_{2}$ is the second Love number\,\cite{Binnington:2009bb,Damour:2012yf}, and $M$ and $R$ are the neutron star mass and radius, respectively. A great virtue of the tidal polarizability is its high sensitivity to the NS radius ($\Lambda\!\sim\!\!R^{5}$) a quantity that has been notoriously difficult to constrain\,\cite{Ozel:2010fw,Steiner:2010fz,Suleimanov:2010th, Guillot:2013wu,Lattimer:2013hma,Heinke:2014xaa,Guillot:2014lla, Ozel:2015fia,Watts:2016uzu,Steiner:2017vmg,Nattila:2017wtj}.

Pictorially, a ``fluffy" neutron star having a large radius is much easier to polarize than the corresponding compact star with the same mass but a smaller radius. This can be achieved by having an EOS with {\it low} pressure at densities near twice nuclear (energy) density, $2\rho_0$. Therefore, if the EOS has {\it too high} a pressure, neutron stars will be too extended and polarizable to be consistent with the gravitational wave data. Remarkably, the tidal polarizability determined from the first BNS merger is already stringent enough to rule out a significant number of previously viable EOSs\,\cite{Abbott:PRL2017} for this very reason.  Furthermore this limit on the deformability has been made even more stringent by requiring both NS to have the same EOS \cite{LIGO_new,DBROWN}, see also \cite{PhysRevX.9.011001}.     

In contrast, radio observations \cite{demorest,2.14Msun} have set an important lower bound on the maximum mass of a neutron star.  In 2010 Demorest et al. found a nearly two solar mass neutron star \cite{demorest} and more recently, measurement of Shapiro delay determined that the millisecond pulsar PSR J0740+6620 has a mass of 2.14$^{+0.10}_{-0.09}M_\odot$ \cite{2.14Msun}.  This large mass NS is supported by an EOS with {\it high} pressure at high densities above $2\rho_0$.  Its observation rules out all EOSs with {\it too low} a pressure at these densities to prevent collapse of the star to a black hole.  

The deformability is most sensitive to the pressure near twice nuclear density $\rho_0$ while the maximum mass depends on the pressure at higher densities.  Therefore a small deformability and a large maximum mass can be simultaneously obtained with an EOS that has a relatively low pressure near $2\rho_0$ and then has a rapidly increasing pressure at higher densities. The tension between small deformability and large maximum mass is manifest in a possible rapid increase in the pressure $p$ with density $\rho$, implying a large speed of sound $c_s =\sqrt{dp/d\rho}$, in units of the speed of light $c$. 

Bedaque and Steiner \cite{Bedaque} and  Moustakidis et al. \cite{PhysRevC.95.045801} have previously discussed the speed of sound in dense matter although these works were done before the deformability observations from GW170817. Alford, Han, and Prakash found that the sound speed likely lies above the conformal limit of $c_s=1/\sqrt{3}$ when investigating generic hybrid stars' EOS, suggesting that quark matter at these densities may be strongly interacting \cite{Alford:2013}. Tews et al. \cite{Tews_2018} as well as Chamel et al. \cite{Chamel:2011aa} also find that the speed of sound in dense matter may be large by considering the maximum mass and radius of a NS. However, X-ray observations of NS radii may be more model dependent than gravitational wave observations of $\Lambda$. More recently, Margaritis et al. claim that the sound speed likely exceeds the conformal limit by studying maximally rotating neutron stars \cite{Margaritis:2019}.

In this paper, we explore a possible tension between the upper limit on the deformability, which favors a soft (low pressure) EOS, and the large limit on the maximum mass, which favors a stiff (high pressure) EOS. This tension may be resolved by considering a large sound speed in the dense matter EOS of a neutron star. In Sec. II, we calculate several EOS by modifying twelve representative EOSs to include a sharp phase transition at high density with a constant sound speed. Han and Steiner have shown that such a transition is effective at reducing the deformability \cite{Han:2019}. In Sec. III, we present our calculations of $\Lambda$ and the sound speed and determine a range of sound speeds which are consistent with the deformability constraints from GW170817. Finally, we discuss the implications of our findings in Sec. IV and conclude in Sec. V.
 
\section{Modified Equations of State}
\subsection{Constant Sound Speed EOS}
We illustrate tensions between the maximum mass of neutron stars and $\Lambda$ by modifying some model EOSs in a simple way to reduce the deformability while maintaining a large maximum mass.  We consider various relativistic and non-relativistic model EOSs at low densities $\rho<\rho_1$, 
\begin{equation}
p(\rho<\rho_1)=p_{\rm model}(\rho)\, .
\label{Eq.model}
\end{equation}
In general we choose $\rho_1=\rho_0=150$ MeV/fm$^3$ to be equal to nuclear density. This may be the lowest density where significant changes in the EOS are still possible without modifying known nuclear structure results.  For example the PREX, PREX II, and CREX experiments, that measure the neutron skin thickness of $^{208}$Pb and $^{48}$Ca, constrain the pressure near $0.66\rho_0$ \cite{PhysRevLett.85.5296,PREX,PhysRevC.85.032501,Horowitz:2014crex}.

\begin{figure}[t]
 \includegraphics[width=1\columnwidth]{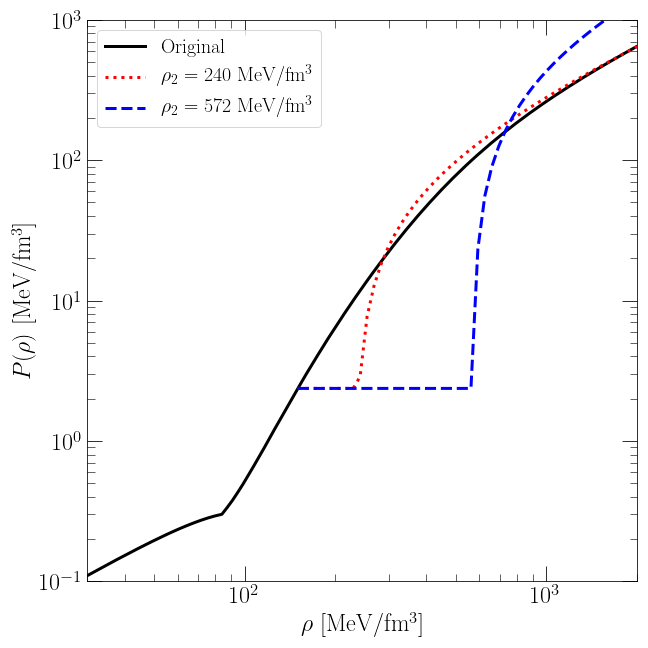}
 \caption{(Color on line) Equation of state --- pressure $p$ versus energy density $\rho$.  A model EOS, in this case the IUFSU relativistic mean field EOS is shown as the solid black line while the dotted red line shows Eq. \ref{Eq.eos} with $\rho_2=240$ MeV/fm$^3$ and $C_s^2=0.365$.  Finally the dashed blue line shows Eq. \ref{Eq.eos} with $\rho_2=572$ MeV/fm$^3$ and $C_s^2=1$.} 
 \label{Fig:eos}
\end{figure}

At higher densities $\rho>\rho_1$, we assume a first order phase transition where the pressure becomes,
\begin{equation}
p(\rho>\rho_1)=p_{\rm model}(\rho_1)+(\rho-\rho_2)C_s^2\Theta(\rho-\rho_2)\, .
\label{Eq.eos}
\end{equation}
Here $\rho_2$, and $C_s^2$ are constants.  We choose $C_s^2$ so that the maximum mass of a NS is $2.1M_\odot$ (consistent with the PSR J0740+6620 observations) and $\Theta(\rho-\rho_2)$ is the Heaviside step function. The high density phase has density $\rho_2$ and a sound speed squared of $C_s^2$ in units of $c^2$.  This forces our EOS at high densities $\rho>\rho_2$ to have a constant speed of sound equal to $c_s=\sqrt{C_s^2}\,$.  Some examples of our EOSs are plotted in Fig. \ref{Fig:eos} using the relativistic IUFSU \cite{Fattoyev:2010mx} model for $p_{\rm model}(\rho)$ . We list all model EOSs used in Table \ref{tab:eos}.

We do not necessarily expect the speed of sound to be constant (independent of density) in the core of a NS. However our choice of a constant sound speed may provide a conservative lower estimate for the maximum speed of sound in dense matter. This constant sound speed model has been a very powerful choice EOS for examining neutron star properties with a generic EOS \cite[e.g. ][]{Alford:2013,Han:2019,Chatziioannou:2019}. If the speed of sound is smaller for some densities, it may be necessary for the speed of sound to be even larger at other densities in order to support the same maximum mass. Our assumption of a sharp first order phase transition with $dp/d\rho=0$ (for $\rho_1<\rho<\rho_2$) is a convenient way to decrease the deformability significantly \cite{Han:2019}. If instead of a phase transition, one has a density region where $dp/d\rho$ is small, we expect qualitatively similar results although with somewhat larger values of $\Lambda$. 

\subsection{Tidal Deformability Calculation}
After one solves the Tolman-Oppenheimer-Volkoff (TOV) equations for the structure of a NS, we then only need to calculate the second Love number $k_2$ in the calculation of $\Lambda$. The second Love number is calculated via:
\begin{multline}
\label{love_num} 
k_2 = \frac{1}{20}\xi^5(1-\xi)^2\Big[(2-y_R)+(y_R-1)\xi\Big]\times \\  \Big\{\Big[(6-3y_R)+\frac{3}{2}(5y_R-8)\xi\Big]\xi+\\
\frac{1}{2}\Big[(13-11y_R+\frac{1}{2}(3y_R-2)\xi+\frac{1}{2}(1+y_R)\xi^2\Big]\xi^2+\\ 3\Big[(2-y_R)+(y_R-1)\xi\Big](1-\xi)^2\ln(1-\xi)\Big\}^{-1}
\end{multline}
Here $\xi$ is the stellar compactness defined as $\xi=2GM/c^2R$ and $y_R=y(R)$ is a dimensionless quantity which is calculated by solving the following nonlinear differential equation:
\begin{equation}
    r\frac{dy}{dr}+y^2+F(r)y+r^2Q(r)=0
    \label{y_ode}
\end{equation}
with the initial condition $y(0)=2$. The functions $F(r)$ and $Q(r)$ are given from:
\begin{eqnarray}
&F(r) = \dfrac{r-4\pi Gr^3(\rho(r)-p(r))}{r-2GM(r)} \\
&Q(r) = \dfrac{4\pi r}{r-2GM(r)}\Big[G\Big(5\rho(r)+9p(r)+\dfrac{\rho(r)+p(r)}{c_s^2(r)}\Big)\nonumber\\
&-\dfrac{6}{4\pi r^2}\Big]- 4\Big[\dfrac{G(M(r)+4\pi r^3p(r))}{r(r-2GM(r))}\Big]^2
\end{eqnarray} in units where $c=1$ \cite{Hinderer:2007mb,piekarewicz:2019}.

Normally one would simply use the results from solving the TOV equations to then integrate Eq. \ref{y_ode} outward to the surface of the star and then solve for $k_2$ with Eq. \ref{love_num}. However, it should be noted that due to the phase transition in the EOS we have implemented, there is a point in which there is a discontinuity in the sound speed when the density sharply goes from $\rho_2\to \rho_1$. Because of this, one must treat this point with special care in the calculation of $y(r)$ as found in refs. \cite{Postnikov:2010yn,Han:2019}. This discontinuity in $y(r)$ can be alleviated by taking the Taylor expansion of $y(r)$ around the radius where the discontinuity takes place, $r_d$. Explicitly this is done via:
\begin{eqnarray}
\lim_{h\to 0}\ y(r_d+h) = y(r_d-h) - 3\frac{\rho_2-\rho_1}{\Bar{\rho}}
\end{eqnarray}
where $\Bar{\rho}$ is $\frac{3M(r_d)}{4\pi r_d^3}$, the average density of the core at $r_d$. For more details see ref. \cite{Postnikov:2010yn}.

To summarize, we generate new EOSs using Eqs. \ref{Eq.model} and \ref{Eq.eos} by picking a model EOS at low density and fixing $\rho_1=150$ MeV/fm$^3$.  Next, a value for $\rho_2$ is chosen and $C_s^2$ adjusted so that the maximum NS mass is $2.1M_\odot$.  All of our EOSs have this maximum mass.  The defomability of a $1.4M_\odot$ NS ($\Lambda_{1.4}$) is calculated as in refs. \cite{DeformabilityCalc,Postnikov:2010yn}. This procedure is repeated for increasing $\rho_2$ values until $C_s^2=1$.
%
 
\begin{figure}[ht]
 \includegraphics[width=1\columnwidth]{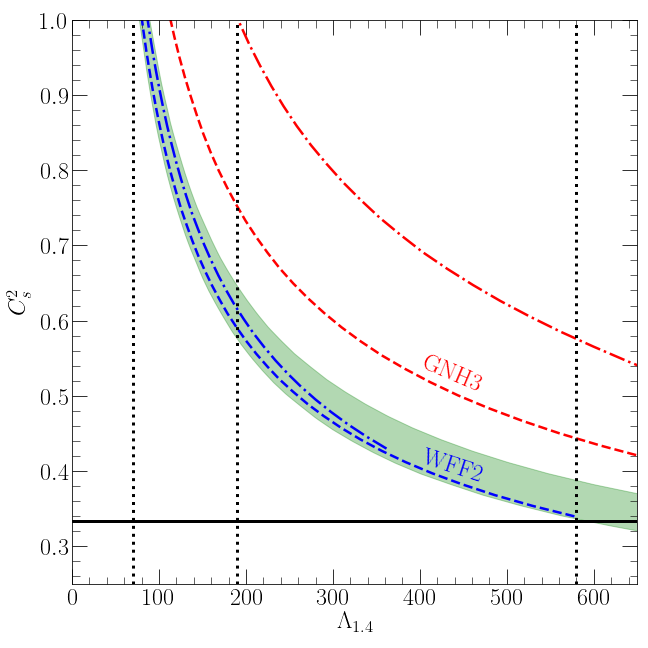}
 \caption{Square of the speed of sound $C_s^2$ versus deformability  $\Lambda_{1.4}$ of a 1.4 $M_\odot$ neutron star.  The shaded region represents the range of sound speeds from the EOSs shown in Table \ref{tab:eos}. The red and blue curves correspond to the $\Lambda_{1.4}$-$C_s^2$ relations with $\rho_1=225$ MeV/fm$^{3}$ (dashed) and with $\rho_1=300$ MeV/fm$^{3}$ (dash-dot). The choice of using the EOSs GNH3 and WFF2 here is to represent the upper and lower-most bounds on sound speed. Also shown by the vertical dotted lines are the LIGO observation $\Lambda_{1.4}=190^{+390}_{-120}$ from GW170817 \cite{LIGO_new}. The solid horizontal line represents where $C_s^2=1/3$.}
 \label{Fig:lambda_cs}
\end{figure}
\section{Results}  
 The resulting values of sound speed squared $C_s^2$ and the deformability $\Lambda_{1.4}$ are shown in Fig. \ref{Fig:lambda_cs}. We find that our $C_s^2$ values depend only slightly on the choice of model low density EOS. Thus, we have a narrow band of $C_s^2$ values as seen in Fig. \ref{Fig:lambda_cs}.  We expect the actual maximum value of the speed of sound, for realistic EOSs with varying sound speeds, to be above this band. 
 
 The $C_s^2$ band rapidly increases as $\Lambda_{1.4}$ decreases. This means, for example, if future GW observations determine $\Lambda_{1.4}$ to be $\approx 190$ (the central value of the LIGO GW170817 observations, $\Lambda_{1.4}=190^{+390}_{-120}$) then Fig. \ref{Fig:lambda_cs} suggests $C_s^2>0.55$. Thus the maximum value of the speed of sound in a neutron star exceeds $0.74c$.  We conclude from Fig. \ref{Fig:lambda_cs} that the maximum speed of sound in a NS is likely large (assuming $\Lambda_{1.4}$ is indeed small). Furthermore, this conclusion is greatly strengthened if future observations further limit $\Lambda_{1.4}$. 
 
 We also partially explore the possibility of a phase transition at densities $\rho_1$  greater than $\rho_0$ in Fig. \ref{Fig:lambda_cs}. We find that $C_s^2$ increases for larger $\rho_1$ as there is more of the nuclear EOS core that makes up the structure of the star.  However, the model dependence is larger than in choosing $\rho_1=\rho_0$, thus the constraints on $C_s^2$ are larger and broader. Additionally, we vary the maximum mass constraint to coincide with the 2$\sigma$ errors on the mass of PSR J0740+6620. Our result that the sound speed is likely large does not change, but the exact bound changes slightly depending on the maximum mass.

\begin{figure}[ht]
 \includegraphics[width=\columnwidth]{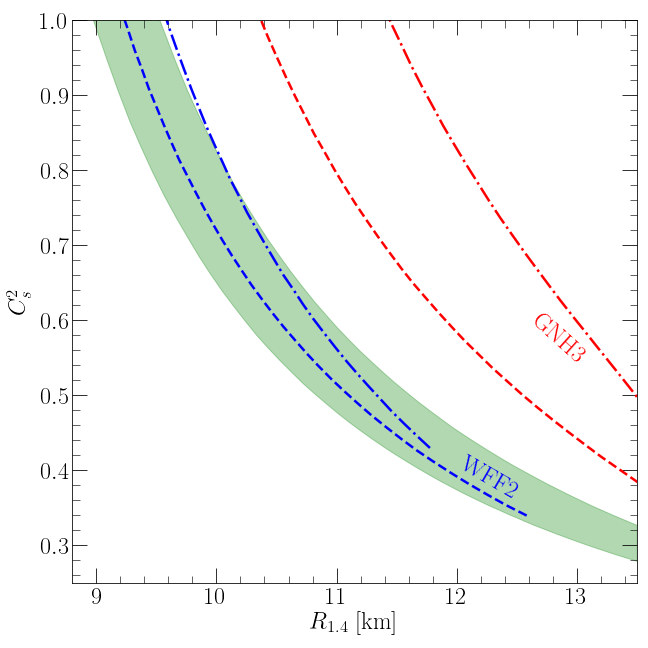}
 \caption{Square of the speed of sound $C_s^2$, in units of $c^2$, versus radius $R_{1.4}$ of a 1.4 $M_\odot$ neutron star.  Like in Fig. \ref{Fig:lambda_cs}, the shaded region represents the full extent of our EOSs. The dashed and dot-dash lines indicate EOSs where $\rho_1=225$ MeV/fm$^3$ and $\rho_1=300$ MeV/fm$^3$ respectively.}
 \label{Fig:radius_cs}
\end{figure}

Figure \ref{Fig:radius_cs} shows $C_s^2$ versus the radius of a $1.4M_\odot$ NS ($R_{1.4}$).  Like Fig. \ref{Fig:lambda_cs} this shows that $C_s^2$ is large if $R_{1.4}$ is small.  However now there is more dependence on the model low density EOS.  One way to think about this extra model dependence is if an extra low density region is added to a NS.  This could increase $R_{1.4}$ without changing $\Lambda_{1.4}$ very much.  Thus $R_{1.4}$ may be more sensitive to exactly how the crust is treated than $\Lambda_{1.4}$ \cite{PhysRevC.99.045802}. 

Figure \ref{Fig:lambda_cs} reveals a minimum value for $\Lambda_{1.4}$ that occurs when $C_s^2=1$.  If an EOS supports a 2.1$M_\odot$ NS, and the EOS does not differ widely from nuclear physics expectations at low densities, then causality requires $\Lambda_{1.4}$ to be larger than this limit.  These minimum deformabilities are provided in Table \ref{tab:eos}.  For all of the EOSs in Table \ref{tab:eos} our absolute lower limit on deformability is $\Lambda_{1.4}\ge 77.35$. This (mostly) EOS independent lower bound on $\Lambda_{1.4}$ may help distinguish neutron stars from black holes.  

\section{Discussion and Implications}
We will now discuss several implications of a large sound speed for the nature of dense matter.  The speed of sound for massless weakly interacting quarks and gluons is $\approx 3^{-1/2}c$.  Asymptotic freedom in QCD implies that this should be the sound speed in the limit of extremely high densities where the interactions between quarks and gluons are expected to be weak.   Our finding that $c_s$ likely exceeds $3^{-1/2}c$ suggests that QCD matter at NS densities is not asymptotically free, but instead strong interactions may be present that increase $c_s$. 

Zel'dovich showed that interactions from the exchange of massive vector particles can increase $c_s$ from $3^{-1/2}c$ to $c$ \cite{Zeldovich}.  Walecka \cite{Walecka} and others, see for example \cite{Serot_Walecka,NIKSIC2011519,PhysRevC.82.055803}, developed relativistic meson baryon field theory models to describe dense matter.  Here nucleons interact via the exchange of scalar (sigma), vector (omega), and possibly other mesons.  Omega exchange naturally leads to these models predicting $c_s\rightarrow c$ at high densities.  Therefore, these models typically predict a high $c_s$ for the dense matter in neutron stars and may easily satisfy our bounds.     

Alternatively, a more traditional picture of nuclear matter has non-relativistic nucleons interacting via two and three nucleon potentials \cite{Naghdi2014}.  Here the two nucleon potential is thought to have a repulsive core at short distances that causes the s-wave phase shifts to become negative at high energies.  In a one-boson exchange picture \cite{MACHLEIDT19871} this hard core has important contributions from the exchange of omega mesons.  Thus the traditional nucleon hard core is likely closely related to the massive vector exchange of relativistic models.

One might expect the nucleon hard cores to give very repulsive contributions and a large sound speed when the cores start to strongly overlap. Interestingly, nuclear matter saturation has proven to be somewhat subtle and likely involves important contributions from three or more body forces \cite{PhysRevLett.47.226,PhysRevC.88.044302}.  Many traditional calculations with only two nucleon forces saturate nuclear matter at too high a density (perhaps twice nuclear density).  This suggests that the hard cores may be somewhat smaller than the distance between nucleons at normal nuclear density and that the hard cores may not overlap strongly until higher densities.   If this is the case, then there may be a higher density, perhaps a few times nuclear density, where the hard cores do interact strongly and this likely would cause a large sound speed.  
For example, the non-relativistic EOS of Akmal, Pandharipande, and Ravenhall (APR) \cite{PhysRevC.58.1804} has a very high sound speed that even exceeds $c$ at very high densities.  This acausal behavior shows that the non-relativistic formalism is incomplete.  Nevertheless, the large sound speed may be qualitatively correct.        

More recently, chiral effective field theory (CEFT) provides a very useful framework to describe neutron rich matter at low densities, see for example \cite{PhysRevC.88.025802}.  Here interactions are expanded in powers of momentum transfer over a chiral scale \cite{PhysRevC.53.2086,sym8040026,Machleidt_2016}.  Long range pion exchanges between nucleons are treated explicitly.  However, any shorter range contributions from the exchange of heavier mesons such as omegas are not calculated explicitly.  Instead their contributions are included with a series of contact terms.  These contact terms may not cleanly resolve the size of any possible nucleon hard cores and thus CEFT may not be able to accurately predict the density where the hard cores first start to strongly overlap.  Indeed the chiral expansion is only expected to converge at low densities.  We concluded that the high sound speed in neutron stars may occur at high densities beyond where CEFT converges.  Nevertheless, the high sound speed observed in neutron stars may provide insight into how the chiral expansion breaks down.  The high sound speed suggests it breaks down in such a way that a series of higher order terms all contribute coherently to enhance the sound speed.   

\begin{table}
   
    \begin{tabular}{c|c|c|c|c}\hline\hline
    EOS &  $C_s^2$ & $R_{1.4}$ & $\rho_2$ & $\Lambda_{min}$  \\
    &\hspace{32 pt} & \hspace{12 pt}[km]\hspace{12 pt} & [MeV/fm$^3$]  & \\\hline
    IUFSU \cite{Fattoyev:2010mx}& 0.588 & 10.58 & 382.98 & 79.48 \\
    FSUGarnet\cite{Chen:2014mza} & 0.589 & 10.60 & 384.36 & 79.87 \\
    FSUGold 2 \cite{Chen:2014sca}& 0.609 & 11.05 & 400.08 & 82.64 \\
    FSUGold 2H \cite{Tolos:2017}& 0.590 & 10.53 & 384.29 & 79.47 \\
    FSUGold 2R \cite{Tolos:2017}& 0.586 & 10.56 & 382.70 & 78.74 \\
    NL3 \cite{Lalazissis:1996rd} & 0.613 & 11.08 & 403.31 & 83.81 \\
    GNH3 \cite{Glendenning:1985} & 0.646 & 10.71 & 419.72 & 89.30 \\
    BSk21 \cite{Potekhin:2013}& 0.583 & 10.41 & 380.09 & 79.29 \\
    WFF2 \cite{Wiringa:1988tp}& 0.576 & 10.36 & 374.56 & 77.35 \\
    MPA1 \cite{Muther:1987} & 0.607 & 10.28 & 393.16 & 82.78 \\
    SLy \cite{Douchin:2000kx}& 0.596 & 10.36 & 387.54 & 80.88 \\
    FPS \cite{Friedman:1981qw}& 0.581 & 10.34 & 378.23 & 78.46 \\
    \hline
    \end{tabular}

     \caption{Comprehensive list of the EOSs used in the text. 
     We report here the values of the square of the sound speed in units of c, $\rho_2$ in $\rm MeV/fm^{3}$, and the radius in km of a $1.4\Msun$ NS at a fixed $\Lambda_{1.4}=190$, the central value given from the LIGO observation of GW170817 \cite{LIGO_new}. We also show the minimum deformabilities of each EOS, determined from where $C_s^2=1$.  We have chosen $C_s^2$ so that all EOSs have a maximum NS mass of $2.1M_\odot$.}
    \label{tab:eos}
\end{table}
Hadronic descriptions, be they relativistic models with vector exchange,  non-relativistic models with hard cores, or CEFT raise a fundamental question.  At NS densities, can one still describe matter in hadronic (meson and baryon) degrees of freedom or must one describe it with quarks and gluons?  Our result of a large sound speed is consistent with meson and baryon degrees of freedom still being applicable at NS densities.  It is not consistent with asymptotically free quarks and gluons. 

However, one may be able to describe NS matter with quark and gluon degrees of freedom, provided one includes strong interactions between the quarks \cite{Baym_2018}.  Thus cold dense matter in a NS may be a strongly interacting quark gluon plasma just as hot dense matter in relativistic heavy ion collisions at RHIC or the LHC is observed to form a strongly interacting quark gluon plasma \cite{QGP_recent,doi:10.1146/annurev-nucl-102212-170540}.  Here strong interactions reduce the quark and gluon mean free paths and lead to a small shear viscosity.   One example of a quark model with strong interactions and a high sound speed is Quarkyonic Matter \cite{PhysRevLett.122.122701}.  Alternatively, strongly interacting quark matter could look very much like strongly interacting hadronic matter, see for example \cite{Alford_2005}.
\section{Conclusion}
In conclusion, gravitational wave observations of the neutron star merger GW170817 limit the deformability of neutron stars and favor low pressure equations of state (EOS).  In contrast, radio observations of the very massive millisecond pulsar PSR J0740+6620 favor high pressure EOS.  Tension between these two observations suggest that the pressure rises rapidly with density and therefore the speed of sound in dense matter may be large --- a significant fraction of the speed of light.  Given these observations, we have used simple model EOSs to place a lower limit on the speed of sound in dense matter.  The speed of sound must be above the green band in Fig. \ref{Fig:lambda_cs}. Our bound becomes even more stringent if future gravitational wave observations further reduce the upper limit on $\Lambda_{1.4}$.  This lower limit on sound speed has important implications for many different hadronic and quark gluon descriptions of dense matter.

\begin{acknowledgments}
This work benefited from discussions at the First Frontiers Summer School supported by the National Science Foundation under Grant No. PHY-1430152 (JINA Center for the Evolution of the Elements).
This material is based upon work supported by the U.S. Department of Energy Office of Science, Office of Nuclear Physics under Awards DE-FG02-87ER40365 (Indiana University) and Number DE-SC0008808 (NUCLEI SciDAC Collaboration). 

\end{acknowledgments}


\end{document}